\documentclass[twocolumn,preprintnumbers,amsmath,amssymb,floatfix,prd]{revtex4}
\usepackage{dcolumn}
\usepackage{ulem}
\usepackage{lipsum}
\usepackage{graphicx}
\usepackage{float}
\begin{document}
\title{Eta Fragmentation Functions Revisited}
\author{Christine A. Aidala}
\email{caidala@umich.edu} 
\affiliation{Department of Physics, University of Michigan, Ann Arbor, Michigan 48109-1040, USA}
\author{Devon A. Loomis}
\email{dloom@umich.edu} 
\affiliation{Department of Physics, University of Michigan, Ann Arbor, Michigan 48109-1040, USA}
\author{Ramiro Tomas Martinez}
\email{rmartinez@df.uba.ar} 
\affiliation{Universidad de Buenos Aires, Facultad de Ciencias Exactas y Naturales, Departamento de F\'{\i}sica 
and IFIBA-CONICET, Ciudad Universitaria, (1428) Buenos Aires, Argentina}
\author{Rodolfo Sassot}
\email{sassot@df.uba.ar} 
\affiliation{Universidad de Buenos Aires, Facultad de Ciencias Exactas y Naturales, Departamento de F\'{\i}sica 
and IFIBA-CONICET, Ciudad Universitaria, (1428) Buenos Aires, Argentina}
\author{Marco Stratmann}
\email{marco.stratmann@uni-tuebingen.de}
\affiliation{Institute for Theoretical Physics, University of T\"ubingen, Auf der Morgenstelle 14, 72076 T\"ubingen, 
Germany}
\begin{abstract}
We revisit the extraction of parton-to-eta meson fragmentation functions at next-to-leading order
accuracy in QCD in the light of the recent hadroproduction measurements in proton-proton collisions 
obtained by the PHENIX, LHCb, and ALICE collaborations. In addition to an increased precision, 
the data explore complementary rapidity ranges and center-of-mass system energies.
The analysis exploits the theoretical scale dependence to ease tensions among the data sets 
at different energies that are potentially caused by QCD corrections beyond the next-to-leading order. 
The resulting set of fragmentation functions yields a consistent description of all available data.
Estimates of uncertainties are obtained with the Monte Carlo replica method.
\end{abstract}

\maketitle

\section{Introduction}
Fragmentation functions (FFs) are evolving in recent years into increasingly precise tools for both 
unveiling the path to confinement of quarks and gluons into hadrons \cite{Field:1976ve}, as well as 
for testing key properties of Quantum Chromodynamics (QCD) such as factorization and universality 
\cite{Collins:1981uw,Collins:1989gx}.
This remarkable improvement is being made possible by vigorous experimental programs pursuing greater 
precision and kinematic coverage, together with achieving milestones in higher order perturbative 
calculations \cite{Mitov:2006ic,Almasy:2011eq,Goyal:2023zdi,Bonino:2024qbh,Czakon:2025yti}, both 
combined in global QCD analyses \cite{Khalek:2021gxf,Borsa:2022vvp}.

Even though FFs are as fundamental as parton distribution functions (PDFs), their development has often been
hindered by the difficulties inherent to obtaining precise single particle inclusive data required for their 
extraction. In addition, the relevant perturbative computations of cross sections with identified
final state hadrons are typically more cumbersome than for fully inclusive quantities or jets.
Nevertheless, FFs offer many interesting and important avenues, now and in the future,  
to fully explore their great phenomenological potential. 
For instance, the possibility of varying fairly straightforwardly the detected final state hadron species,
allows one to investigate and parametrize the still poorly understood hadron mass, parton flavor, 
and perhaps even spatial dependence of the nonperturbative hadronization process. 
Observed hadrons can also be used as tools to deepen our understanding of 
proton-nucleus and heavy-ion collisions by studying modifications of their yields due
to medium-induced effects, see, e.g., \cite{Doradau:2024wli}.

Fifteen years ago, the first global QCD analysis of eta meson ($\eta$) FFs at 
next-to-leading order (NLO) accuracy was presented in reference \cite{Aidala:2010bn} (AESSS). Back then it
combined world data on single inclusive electron-positron annihilation (SIA) experiments 
\cite{ref:hrs,ref:mark2,ref:jade1,ref:jade2,ref:cello,ref:aleph1,ref:aleph2,ref:aleph3,ref:l31,ref:l32,
ref:opal,ref:babar} with the
proton-proton collision data \cite{Adler:2007phenix,Adare:2011phenix}
available at that time, finding a good agreement within uncertainties. 
Since then, other groups have also addressed $\eta$ fragmentation. 
In ref.~\cite{Li:2024etc} only SIA data were included, which on their own do not 
suffice to determine all aspects of parton-to-$\eta$ FFs, in particular, to verify 
their important universality property assumed in perturbative QCD.
The fit described in \cite{Gao:2025bko} (NPC23) considers in addition results
of proton-proton cross section ratios to pions, which potentially obscure, however,
issues with the individual absolute normalizations as we shall discuss later.
Clearly, only a global analysis of cross section data can fully exploit the discriminating power 
of results from various different experiments in a fit of FFs and this is what
we are going to pursue in what follows.

In the past couple of years, the ALICE 
\cite{Abelev:2012alice,Acharya:2017alice,Acharya:2018alice,Acharya:2021alice} and the 
LHCb \cite{Aaij:2024lhcb} experiments at the CERN Large Hadron Collider (LHC) 
have explored hadroproduction of $\eta$ mesons as a function of their transverse momentum $p_T$
for center-of-mass system (c.m.s.) energies $\sqrt{S}$ ranging from 2.7 to $13\,\mathrm{TeV}$ 
at both central and forward rapidities $y$. 
All LHC results are at odds with theoretical estimates of the corresponding $\eta$ hadroproduction
cross sections computed at NLO accuracy with the AESSS FFs of reference \cite{Aidala:2010bn}
which grossly overshoot the data as we shall show in the figures below.
This rather large discrepancy could, in principle, point to shortcomings with either the determination of the FFs 
by AESSS, which solely used data at much lower c.m.s.\ energies, or the 
perturbative approximation to NLO or, most likely, with a combination of both.
In this respect it should be noted that a similar disagreement between data and theory has also been observed 
for single inclusive pion and charged hadron production since the start of the LHC. 
The PHENIX experiment located at the Relativistic Heavy Ion Collider (RHIC) 
at Brookhaven National Laboratory (BNL), whose midrapidity results 
obtained at $\sqrt{S}=200\,\mathrm{GeV}$ \cite{Adler:2007phenix,Adare:2011phenix} 
were the only source of $pp$ data in the AESSS fit,
has complemented their suite of $\eta$ production results
by data in a forward rapidity interval \cite{Adare:2014phenix} and, most recently,
also by mid- and forward rapidity data at a higher c.m.s.\ energy of
$\sqrt{S}=510\,\mathrm{GeV}$ \cite{Abdulameer:2025phenix}. 

In updated global extractions of pion and charged hadron FFs at 
NLO accuracy \cite{Borsa:2021ran,Borsa:2023zxk} 
based on the latest LHC results it has been found that the above mentioned discrepancy
between data at different c.m.s.\ energies and theory calculations persists. 
It has been noted, however, 
in \cite{Borsa:2021ran,Borsa:2023zxk} that the 
theoretical description of the global set of data 
can be significantly improved by fully exploiting the large
uncertainties due to the truncation of the perturbative series at NLO accuracy as
is reflected by the variations in cross section estimates
induced by different choices of the renormalization, initial-, and final-
state factorization scales, $\mu_{R,FI,FF}$.
Therefore, an extra fit parameter $\kappa(\sqrt{S})$ was introduced
in Refs.~\cite{Borsa:2021ran,Borsa:2023zxk} 
to account for, in the simplest possible approximation, a possible c.m.s.\ energy dependence 
of the missing higher order QCD corrections by rescaling $\mu_{R,FI,FF}$ 
with $\kappa(\sqrt{S})$ in the global fit. 

In a recent, pioneering next-to-next-to-leading order (NNLO) calculation 
of identified hadron production in $pp$ collisions
\cite{Czakon:2025yti} it was found that the NNLO QCD
corrections are indeed significant, in particular, at smaller
c.m.s.\ energies and at small to moderate $p_T$ of the observed hadron.
Unfortunately, the NNLO results cannot be readily implemented into a global
analysis framework for the time being due to their numerical complexity. 

Nevertheless, we think that an update of the AESSS analysis of parton-to-$\eta$ meson FFs
is long overdue in view of the wealth of new experimental results in a vast
kinematic range from both CERN-LHC and BNL-RHIC. 
In preparation for a future global QCD analysis at NNLO accuracy, 
we hence proceed along the lines of Refs.~\cite{Borsa:2021ran,Borsa:2023zxk} to assess 
to what extent $\eta$ meson production can be described at NLO accuracy. 

To estimate residual uncertainties of the extracted $\eta$ FFs we switch from the
rigorous but cumbersome method of Lagrange multipliers used by AESSS \cite{Aidala:2010bn}
to Monte Carlo replicas. The latter method,
which does not require any assumptions about a particular tolerance criterion,
was also adopted in \cite{Borsa:2021ran,Borsa:2023zxk} and greatly
facilitates the propagation of uncertainties to any observable depending on
our newly extracted $\eta$ FFs.

The remainder of the paper is organized as follows: Next, we give a brief outline of
the setup of our global analysis framework and the sets of data used in the fit.
Section III contains a detailed discussion of the results and uncertainties.
We conclude in Section IV.
 
\section{Setup and data sets}
Since the methodology we implement to obtain the $\eta$ FFs from a global analysis at NLO
accuracy follows closely the steps outlined in Refs.~\cite{Borsa:2021ran,Borsa:2023zxk}, we can
be very brief in this section and refer the reader to the literature.
In the following, we shall focus on the peculiarities of $\eta$ mesons  
and the main features of the newly included data sets.

In general, data on identified $\eta$ mesons are less abundant than 
those for charged and neutral pions or kaons and, hence, less precise.
As was already pointed out in the original AESSS analysis \cite{Aidala:2010bn},
the most crucial obstacle in the extraction of parton-to-$\eta$ FFs is,
however, the complete lack of experimental information from semi-inclusive 
deep inelastic scattering (SIDIS). This deprives us of important constraints on 
the flavor separation of light (anti)quarks fragmenting into $\eta$ mesons.
Unlike for pions and kaons, there are, in addition, no flavor-tagged SIA data
which usually play an important role in constraining the charm- and bottom-to-hadron
FFs in a global fit \cite{Borsa:2021ran,Borsa:2023zxk}. 
In both cases we have to resort to assumptions, which introduce a certain
bias to the fit that is hard to quantify. 

Hence, instead of extracting the FFs for the light quarks and antiquarks
individually from data, as for pions and kaons, we start by parametrizing
them under the \textit{assumption} that light $u$ and $d$ (anti)quark FFs 
are all equal and that the $s=\bar{s}$ FFs are proportional to them.
I.e., at our input scale of $\mu_0=1\,\mathrm{GeV}$ we set
\begin{equation}
\label{eq:quarks}
D^{\eta}_u=D^{\eta}_{\bar{u}}=D^{\eta}_d=D^{\eta}_{\bar{d}}
\;\;\text{and}\;\;
D^{\eta}_s=D^{\eta}_{\bar{s}} = N_s\,D^{\eta}_u\;.
\end{equation}
Next, we use the same flexible functional form as in previous analyses 
with five free fit parameters,
\begin{eqnarray}
\label{eq:ansatz}
D_{i}^{\eta}(z,\mu_0)  =  
N_{i} \,z^{\alpha_{i}}(1-z)^{\beta_{i}} [1+\gamma_{i} (1-z)^{\delta_{i}}]\;,
\end{eqnarray}
where $i=u$ and $z$ denotes the fraction of the four momentum of the parton taken by the 
$\eta$ meson.
This ansatz differs from the AESSS fit \cite{Aidala:2010bn} in which it
was assumed that all light quarks, $u$, $d$, and $s$, fragment equally into
$\eta$ mesons. Either way, such assumptions can be easily motivated from the wave function 
of the $\eta$ meson as was discussed in detail in Ref.~\cite{Aidala:2010bn}.
Introducing a second independent parametrization like in Eq.~(\ref{eq:ansatz}) for $i=s$ to
\textit{fully} discriminate between strange and nonstrange FFs does not improve the quality
of our fit any further. This is because the quark FFs are
mainly determined in the global fit by the same suite of SIA data as in the AESSS analysis.
They are only rather mildly constrained by the newly added wealth of hadroproduction data
which tend to favor a different normalization $N_s$ in Eq.~(\ref{eq:quarks}) relative
to the $u$ and $d$ (anti)quarks.

The latter hadroproduction data are solely responsible for constraining the gluon-to-eta FF $D_g^{\eta}$.
Of course, there is a cross-talk between (anti)quark and gluon FFs due to scaling
violations which are encoded in the timelike evolution equations. In this way, 
the experimentally observed energy dependence of the SIA data for any given $z$ 
indirectly constrains also the gluon FF in a global analysis. 
However, unlike for PDFs, where scaling violations are a powerful tool to determine
the gluon density at small momentum fractions, their impact in a fit of FFs 
turns out to be very limited in the kinematic range relevant for SIA data.

At the time of the AESSS analysis, only PHENIX data \cite{Adler:2007phenix,Adare:2011phenix}
at a single c.m.s.\ energy of $\sqrt{S}=200\,\mathrm{GeV}$ were available and
setting $\gamma_{g}=0$ in the ansatz Eq.~(\ref{eq:ansatz}) for $D_g^{\eta}(z,\mu_0)$
turned out to be sufficient for a good fit \cite{Aidala:2010bn}. 
The present, much improved suite of hadroproduction data, 
now covering c.m.s.\ energies $\sqrt{S}$ up to $13\,\mathrm{TeV}$, 
allows us to relax this assumption.

For the fragmentation of heavy charm and bottom quarks into $\eta$ mesons, 
we follow a similar strategy as was implemented in \cite{Aidala:2010bn}.
To this end, we link their functional form to the one for charm and bottom FFs into 
residual hadrons as obtained in Ref.~\cite{deFlorian:2007ekg}, 
but, in addition, we allow for different normalizations $N_{c,b}$ and 
a modified low-$z$ behavior $\alpha_{c,b}$ to be constrained by the fit to data:
\begin{eqnarray}
\label{eq:ansatz-hq}
D_{c}^{\eta}(z,m_c) &=& D_{\bar{c}}^{\eta}(z,m_c) = N_c \,z^{\alpha_c}D_{c}^{res}(z,m_c)\;,
\nonumber \\
D_{b}^{\eta}(z,m_b) &=& D_{\bar{b}}^{\eta}(z,m_b) = N_b \,z^{\alpha_b}D_{b}^{res}(z,m_b)\;.
\end{eqnarray}
The parameters specifying the $D_{c,b}^{res}$ can be found in Tab.~III of Ref.~\cite{deFlorian:2007ekg}.
The FFs in Eq.~(\ref{eq:ansatz-hq}) are included discontinuously as massless
contributions in the scale evolution of the FFs above their $\overline{\mathrm{MS}}$ scheme
thresholds $\mu=m_{c,b}$ with $m_{c}=1.43\,\mathrm{GeV}$ and $m_{b}=4.3\,\mathrm{GeV}$ 
denoting the mass of the charm and bottom quark, respectively.

In total, the free fit parameters introduced in Eqs.~(\ref{eq:ansatz}) and (\ref{eq:ansatz-hq})
to describe the FFs of (anti)quarks and gluons into $\eta$ mesons add up to 15.
They are determined by a standard $\chi^2$ minimization for $N=349$ available 
data points for SIA and hadroproduction.

\begin{figure}[thb!]
    \centering
    \includegraphics[width=1\linewidth]{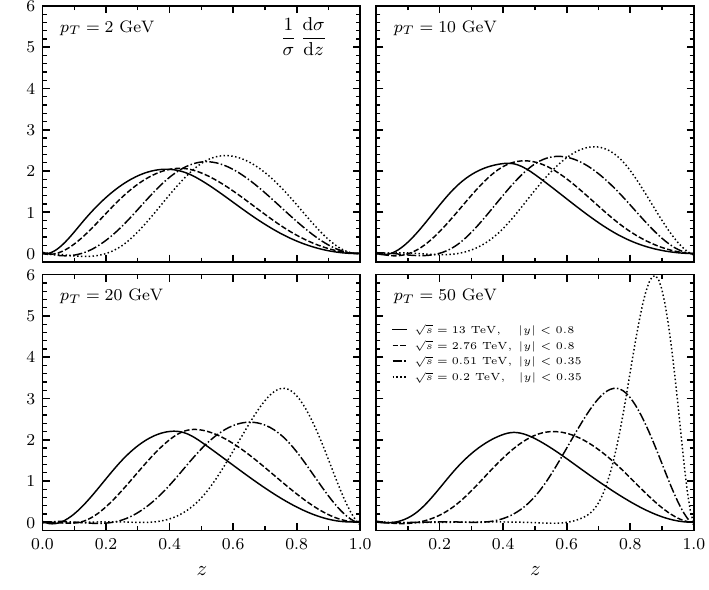}
    \vspace*{-0.4cm}
    \caption{Scaled $z$-differential hadroproduction cross section at midrapidity for four 
    different bins in $p_T$ and for four different c.m.s.\ energies $\sqrt{S}$; see text.}
    \label{fig:z_dist}
\end{figure} 
For the first time in a global QCD fit of $\eta$ meson FFs, we directly use LHC $\eta$ meson 
hadroproduction data from ALICE at $\sqrt{S} = 2.76,\, 7,\, 8$, and $13\,\mathrm{TeV}$
\cite{Abelev:2012alice,Acharya:2017alice,Acharya:2018alice,Acharya:2021alice} and LHCb 
at $\sqrt{S} = 5.06$ and $13\,\mathrm{TeV}$ \cite{Aaij:2024lhcb}. 
The high energy LHC data are supplemented by lower energy $\eta$ meson production data from PHENIX, 
which, for the first time, include central (forward) rapidity measurements 
at $\sqrt{S} = 510\, (500)\,\mathrm{GeV}$
\cite{Abdulameer:2025phenix} and a forward rapidity cross section at $\sqrt{S} = 200\,\mathrm{GeV}$  
\cite{Adare:2014phenix}. These new PHENIX data accompany the earlier 
midrapidity $\sqrt{S} = 200\,\mathrm{GeV}$ results \cite{Adler:2007phenix,Adare:2011phenix} 
that were used already in the AESSS fit \cite{Aidala:2010bn}. 
The forward rapidity results from LHCb and PHENIX add novel, complementary information
to our analysis as they probe a different mixture of partonic hard scattering channels 
than the data collected at central rapidities. While midrapidity $pp$ cross sections
are heavily dominated by gluon-gluon scattering, quark-gluon subprocesses play an
equally important role at forward rapidities. 

To illustrate the much improved constraining power of the experimental inputs from the LHC and RHIC,
we show in Figure~\ref{fig:z_dist} the $z$-differential
hadroproduction cross section at midrapidity scaled by the integrated cross section
in four bins of $p_T$. Hence, each curve gives a direct indication of the range in $z$ 
which is predominantly probed for a given c.m.s.\ energy $\sqrt{S}$.
As can be seen, the larger the ratio $p_T/\sqrt{S}$ the more the peak of the 
$z$-distribution moves to larger values, i.e., for any given fixed $p_T$, experiments at
a lower $\sqrt{S}$ probe larger values of $z$ than experiments at a higher $\sqrt{S}$. 
This is where the PHENIX data add most to the very precise data coming from the LHC. 
Figure~\ref{fig:z_dist} also shows that the z-distributions are rather broad
and have significant overlap with each other for different values of $\sqrt{S}$ (and $p_T$)
which in turn implies that our global analysis will easily reveal any tensions between the
different experimental results.

\begin{figure*}[th!]
    \centering
    \includegraphics[width=0.6\linewidth]{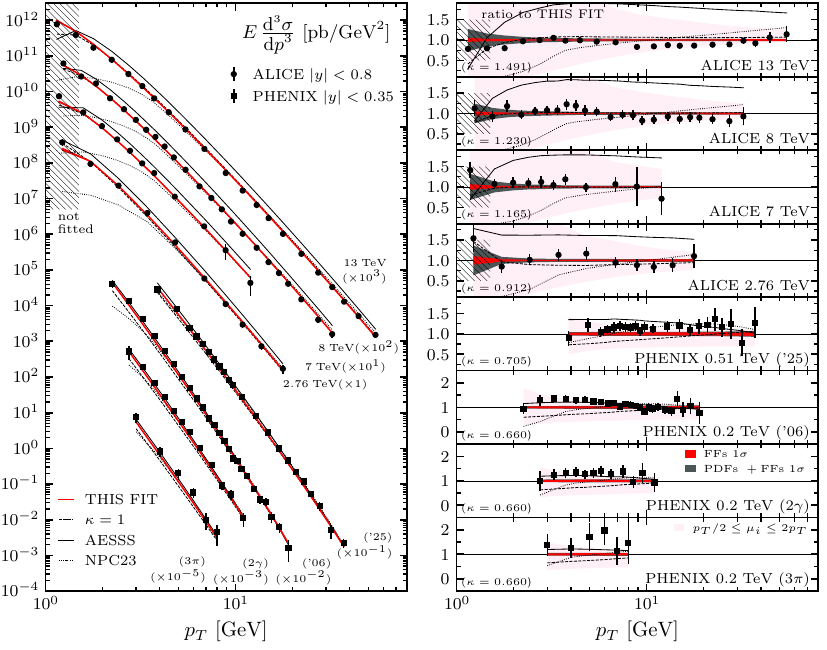}
    \caption{Left-hand-side: comparison of our new NLO results (red lines, labeled as ``THIS FIT'') 
    with the $p_T$-differential midrapidity hadroproduction data from ALICE 
    \cite{Abelev:2012alice,Acharya:2017alice,Acharya:2018alice,Acharya:2021alice} and PHENIX 
    \cite{Adler:2007phenix,Adare:2011phenix,Abdulameer:2025phenix} at various different $\sqrt{S}$.
    Also shown are the results based on a fit using a conventional choice of scales
    $(\kappa=1)$ and by utilizing the AESSS \cite{Aidala:2010bn} and NPC23 \cite{Gao:2025bko} 
    sets of FFs.
    The panels of the right-hand-side show the corresponding ratio to our new fit (``THIS FIT'')
    for each set of data. The shaded bands illustrate various sources of uncertainties,
    see text. Note that data with $p_T<1.5\,\mathrm{GeV}$ are not included in the fit as
    is indicated by the hatched areas.}
    \label{fig:mid}
\end{figure*}

In addition to the wide range of hadroproduction data, we utilize essentially the same
suite of SIA measurements 
\cite{ref:hrs,ref:mark2,ref:jade1,ref:jade2,ref:cello,ref:aleph1,ref:aleph2,ref:aleph3,ref:l31,ref:l32,
ref:opal} that was included in the AESSS fit \cite{Aidala:2010bn},
covering electron-positron c.m.s.\ energies from $\sqrt{S}= 10.54\,\mathrm{GeV}$ 
up to the $Z$ boson resonance at $91.2\,\mathrm{GeV}$.
Here we only swap the preliminary and actually never published BaBar data \cite{ref:babar}
used in AESSS for the new high precision SIA measurement from Belle \cite{Seidl:Belle2025}.
As was discussed in some detail in \cite{Aidala:2010bn}, we convert, if needed, all SIA data to the
usual scaling variable $z$ and impose the standard cut $z>0.1$.
We note that we refrain from using the very low c.m.s.\ energy data from BES-III \cite{ref:besiii}
which are strongly contaminated by unwanted higher twist effects \cite{Li:2024etc}.
Likewise, we do not include $\eta$ meson production data from fixed target, low $\sqrt{S}$ 
hadron-hadron collision experiments, like Fermilab E706 \cite{FermilabE706:2002wtp}
where it is known that theoretical calculations at NLO accuracy cannot
describe the data without including all-order QCD resummations of threshold logarithms
\cite{Hinderer:2018nkb}. As in Refs.~\cite{Borsa:2021ran,Borsa:2023zxk} we apply a general
cut $p_T>1.5\,\mathrm{GeV}$ also on the hadroproduction data from high energy $pp$ colliders to ensure
the applicability to perturbative QCD. Data below that cut are shown in the figures
but are not used in our fit.

The data sets used in this analysis are also summarized in 
Table~\ref{tab:exppiontab} in the next section.

\section{Results}
%
In this section we present the results of our updated global analysis of parton-to-$\eta$ meson
FFs at NLO accuracy based on the framework and data sets outlined in Sec.~II.
In what follows, all $pp$ cross sections are computed with the MSHT20 set of PDFs 
\cite{Bailey:2020ooq} and the corresponding value for the strong coupling at NLO accuracy. 
Any other recent set of PDFs will lead to very similar results within the uncertainties of the fit. 

As was mentioned already in the Introduction, we follow Ref.~\cite{Borsa:2021ran,Borsa:2023zxk}
and allow for extra fit parameters $\kappa(\sqrt{S})\in[1/2,2]$ as simple 
proxies for missing higher order corrections in the hadroproduction and SIA cross sections.
The renormalization, initial and final state factorization scales are then
simultaneously set to $\mu_{R,FI,FF} = \kappa(\sqrt{S})\,p_T$, 
i.e., $\kappa=1$ refers to the conventional choice $\mu_{R,FI,FF} = p_T$ in hadroproduction.
Likewise, in case of SIA we set $\mu_{R,FF} = \kappa(\sqrt{S})\,\sqrt{S}$.
In corresponding global analyses of kaons, protons, antiprotons, and inclusive charged hadrons
\cite{Borsa:2023zxk} it was found that the values of $\kappa(\sqrt{S})$ determined 
in the fit of pion FFs \cite{Borsa:2021ran} yielded very good descriptions of the data 
\textit{without} refitting them. Fortunately, it turns out that this is also the case 
for $\eta$ mesons. The quoted values of $\kappa(\sqrt{S})$ therefore 
refer to those given in Ref.~\cite{Borsa:2021ran}.

\begin{figure*}[th!]
    \centering
    \includegraphics[width=0.9\linewidth]{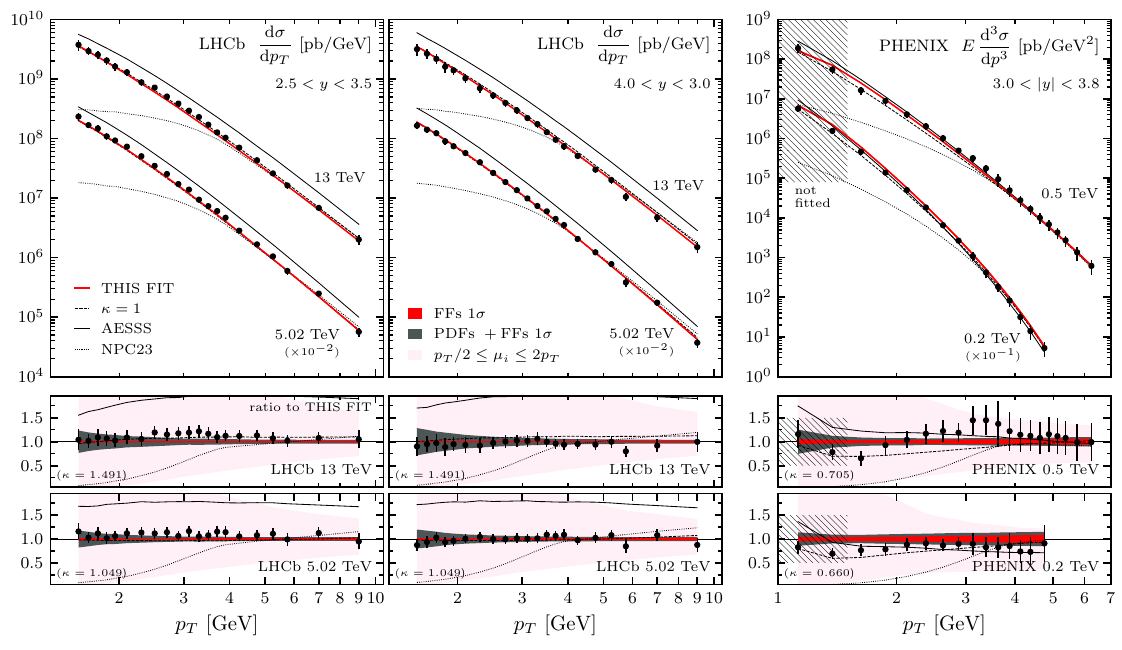}
    \caption{Similar as Figure~\ref{fig:mid} but now for the forward rapidity data from
    from LHCb \cite{Aaij:2024lhcb} and PHENIX \cite{Adare:2014phenix,Abdulameer:2025phenix}.
    The ratios to our new best fit for each set of data can be now found in the panels in the
    lower two rows.}
    \label{fig:fwd_bwd}
\end{figure*}
Uncertainty estimates for our FFs are obtained by a Monte Carlo sampling technique
following the precedure outlined in refs.~\cite{Borsa:2021ran,Borsa:2023zxk}.
In a first step, this amounts to producing 300 statistically equivalent 
replicas of the data sets included in the fit by smearing each data point
within its quoted experimental uncertainty assuming a Gaussian distribution,
see, e.g., Ref.~\cite{DeFlorian:2019xxt}. 
Next, a fit to each of the replicas of the data is performed. The resulting 300
sets of FFs are then expected to inherit the statistical properties 
of the original data sets such that a mean (best) fit and its 1-$\sigma$ standard
deviation can be defined statistically without invoking any tolerance criterion
to judge the quality of the fit.

In Figure~\ref{fig:mid} we compare various theoretical calculations at NLO accuracy with 
the $p_T$-differential midrapidity hadroproduction data from ALICE 
\cite{Abelev:2012alice,Acharya:2017alice,Acharya:2018alice,Acharya:2021alice} and PHENIX 
\cite{Adler:2007phenix,Adare:2011phenix,Abdulameer:2025phenix}. 
The left-hand-side (l.h.s.) of the figure shows the cross sections at various $\sqrt{S}$ 
and the panels on the right-hand-side (r.h.s.) offer a more detailed account of the
quality of the theoretical results by giving 
the corresponding ratio to our new optimum fit (labeled as ``THIS FIT'') for each set of data.

First, one notices that the original AESSS set of FFs still delivers a fairly good description
of the new PHENIX data at $\sqrt{S}=510\,\mathrm{GeV}$ \cite{Abdulameer:2025phenix} which
is, perhaps, not such a surprise as it was designed to fit the corresponding PHENIX data 
at $\sqrt{S}=200\,\mathrm{GeV}$ \cite{Adler:2007phenix,Adare:2011phenix}.
Yet, the already visible trend to slightly overshoot the data at $\sqrt{S}=510\,\mathrm{GeV}$
quickly accelerates with increasing c.m.s.\ energy and the agreement with the ALICE
data \cite{Abelev:2012alice,Acharya:2017alice,Acharya:2018alice,Acharya:2021alice} is 
in general very poor.

Another important observation can be made by comparing next our new fit for the conventional choice
of scales ($\kappa =1$) with the data for the $\eta$ meson invariant cross section. Now the LHC data
at the highest c.m.s.\ energies are nicely described but the agreement deteriorates for
the lower energy PHENIX data, where the fit visibly undershoots the data.
Clearly, the global fit is steered by the more numerous and more precise suite of LHC data in this
case.

The very same apparent tension between LHC data in the TeV-region and hadroproduction results
obtained at lower c.m.s.\ energies has plagued global analyses of other hadron species in 
recent years \cite{Borsa:2021ran,Borsa:2023zxk}. 
Assuming that there is no \textit{real} tension among the experimental results, the problem
must reside in the theoretical framework, more specifically, in the truncation of the
perturbative series at NLO accuracy. The relevance of missing higher order corrections
is usually judged by exploring the scale uncertainties inherent to the NLO results. 
The lightly shaded bands in the panels 
on the r.h.s.\ of Fig.~\ref{fig:mid} illustrate the scale ambiguity
relative to the $\kappa =1$ fit by independently varying $\mu_R$, $\mu_{FI}$, and
$\mu_{FF}$ (``27-point variation'')
by the usual factor of two up and down around the default choice $p_T$.
Clearly, these uncertainties outrival the experimental errors and are
most pronounced towards small values of the hard scale $p_T$ where the use
of perturbative methods reaches its limits.

Following refs.~\cite{Borsa:2021ran,Borsa:2023zxk}, we exploit the freedom in the
choice of scales by adopting the same energy dependent scale factors $\kappa(\sqrt{S})$
in an attempt to arrive at a good global description of all data, i.e., our new 
optimum fit (red lines in Fig.~\ref{fig:mid}).
Of course, this can be only viewed as a simple-minded proxy for the full NNLO corrections 
that have been computed very recently \cite{Czakon:2025yti} but due to their numerical
complexity still lack an appropriate implementation into a global analysis framework.
Hopefully, in the not-too-distant future global analyses of FFs can be elevated to full
NNLO accuracy. 
As can be seen, allowing for energy-dependent scales yields the desired
good agreement to data across the entire range of c.m.s.\ energies and down to
unexpectedly small values of $p_T \simeq \mathcal{O}(1-2\,\mathrm{GeV})$.
The differences to the $\kappa=1$ fit are most pronounced at low $p_T$ and
for smaller $\sqrt{S}$ and fade out for most of the LHC data.

For completeness, Fig.~\ref{fig:mid} also shows the results obtained with the NPC23 set 
of $\eta$ FFs \cite{Gao:2025bko} (dotted lines) that uses essentially the same 
suite of SIA data as in our fit but utilizes only hadroproduction data 
for the $\eta/\pi$ cross section ratio for $p_T>4\,\mathrm{GeV}$.
Below this cut, the fit fails to describe the data by an order of magnitude. Otherwise
it delivers a fair but not optimal description of the data as can be best inferred from
the panels on the r.h.s.\ of Fig.~\ref{fig:mid}.

Finally, the two dark shaded bands (black and red labeled as ``PDFs + FFs $1\sigma$''
and ``FFs $1\sigma$'', respectively) should give an idea of the residual uncertainties 
of our extracted FFs with and without including additional uncertainties of the 
PDFs. In the case of our FFs they are determined statistically by computing the 
$68\%$ confidendence level (C.L.) interval ($1$-$\sigma$ standard deviation) from our 300 replicas.
Likewise, the PDF uncertainties refer to the $68\%$ C.L.\ bands quoted in the
MSHT20 analysis \cite{Bailey:2020ooq}. 
As can be seen, these ambiguities, which just propagate the experimental uncertainties,
are in general significantly smaller than the theoretical uncertainties
due to the choice of scales (light shaded bands).

The forward (and in case of PHENIX also backward) rapidity data shown in 
Figure~\ref{fig:fwd_bwd} exhibit the same pattern of (dis)agreement as we have just discussed
in case of hadroproduction at midrapidity.
On the one hand, the old AESSS fit gives a reasonable description of 
the PHENIX data \cite{Adare:2014phenix,Abdulameer:2025phenix} except for
$p_T\lesssim 2\,\mathrm{GeV}$ but again fails badly for the LHCb data \cite{Aaij:2024lhcb} 
and, on the other hand, the $\kappa=1$ fit does well for LHCb but not for PHENIX. 
The scale ambiguities in this more extreme kinematic regime of forward rapidities, again 
indicated by the light shaded bands, are now even more pronounced 
than in Fig.~\ref{fig:mid}.
The agreement between our best fit based on energy dependent scales (red lines) and
essentially all the data shown in Fig.~\ref{fig:fwd_bwd} is astonishingly good given
the fact that the true NNLO corrections will exhibit a highly nontrivial dependence on 
$\sqrt{S}$, $p_T$ and also rapidity $y$ and not just only on the c.m.s.\ energy as
we have assumed throughout.

\begin{figure}[thb!]
    \centering
    \includegraphics[width=0.75\linewidth]{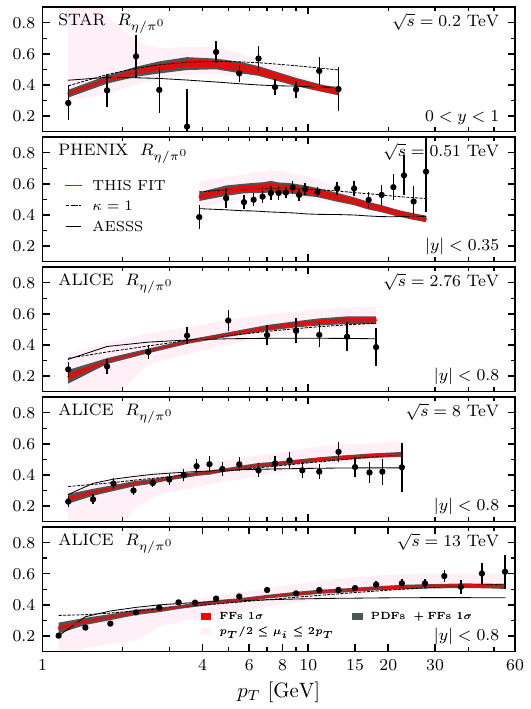}
    \caption{NLO calculations of the $\eta/\pi^0$ cross section ratios
    for our new best and $\kappa=1$ fits and the AESSS set of FFs
    compared to data from STAR, ALICE, and PHENIX \cite{STAR:2009qzv,Abelev:2012alice,Acharya:2017alice,Acharya:2018alice,
    Acharya:2021alice,Abdulameer:2025phenix}. As in Figs.~\ref{fig:mid} and \ref{fig:fwd_bwd}
    the shaded bands illustrate our estimates of uncertainties related to scale variations,
    the extracted FFs, and the MSHT20 set of PDFs \cite{Bailey:2020ooq}.}
    \label{fig:RATIOS}
\end{figure}
Although we make no use of existing data on the eta-to-pion cross section ratios 
$R_{\eta/\pi^{0}}$
\cite{STAR:2009qzv,Abelev:2012alice,Acharya:2017alice,Acharya:2018alice,
Acharya:2021alice,Abdulameer:2025phenix} in our global analysis,
as they are fully correlated with the experimental information 
from the $\eta$ yields that is already included in our fit,
we present a comparison with NLO calculations in Figure~\ref{fig:RATIOS}.
It should be noted that we use the latest pion fragmentation 
functions from \cite{Borsa:2021ran} in the computation of the $\eta/\pi^0$ ratio 
with our present fits, with and without modifications of the scales, but,
for consistency, the older set of pion FFs from \cite{deFlorian:2007aj}
in calculations based on the AESSS $\eta$ FFs.
We should mention that the STAR collaboration has only presented measurements \cite{STAR:2009qzv}
of $R_{\eta/\pi^{0}}$ but not for the $\eta$ hadroproduction cross section.
Because of that and the limited precision of the data on $R_{\eta/\pi^{0}}$
we refrain from using these results in our fit.

\begin{figure*}[thb!]
    \centering
    \includegraphics[width=0.75\textwidth]{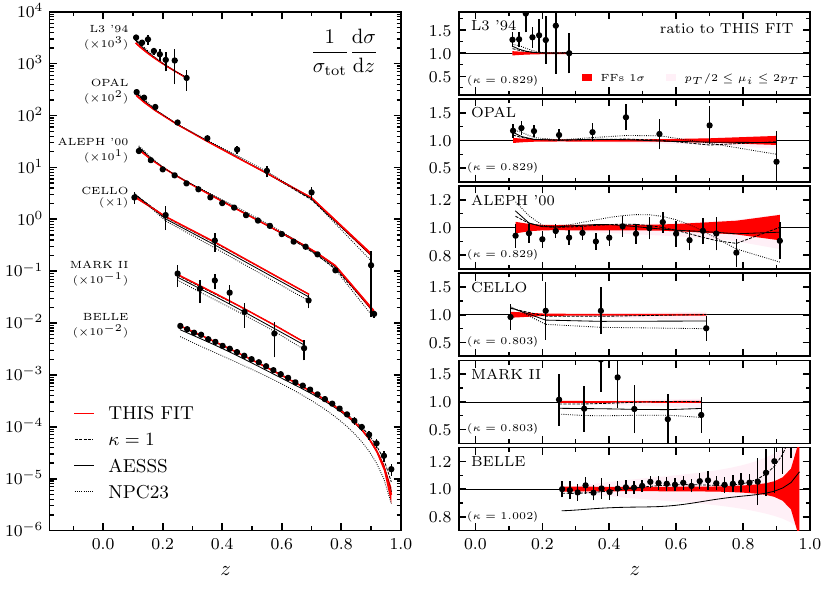}
    \caption{Similar to Fig.~\ref{fig:mid} but now for 
    a representative set of $\eta$ meson production data in SIA, see text.
    }
    \label{fig:SIA}
\end{figure*}
Within the uncertainties of the data, both the best and the $\kappa=1$ fit describe
all the eta-to-pion cross section ratios $R_{\eta/\pi^{0}}$ very well, while the 
results obtained with the AESSS set represent the trend of the data less
favorably. These results have to be taken with a grain of salt though. 
Most importantly, ratios often tend to hide serious problems with the absolute normalizations
of the cross sections entering the ratio because, for instance, 
missing higher order corrections affect the numerator and denominator in a similar way
and thus tend to cancel despite being crucial for a good theoretical description.
This also happens here. The inadequacy of the $\kappa =1$ set in 
reproducing the PHENIX data in Fig.~\ref{fig:mid} is completely obscured in
the corresponding ratio $R_{\eta/\pi^{0}}$ because a $\kappa=1$ fit 
fails in a similiar fashion also for pions, see Ref.~\cite{Borsa:2021ran}.
This is the main reason why we refrain from using cross section ratios in a global analysis.

In Figure~\ref{fig:SIA} we present a representative sample of SIA data sets included in our fit 
\cite{ref:hrs,ref:mark2,ref:jade1,ref:jade2,ref:cello,ref:aleph1,ref:aleph2,ref:aleph3,ref:l31,ref:l32,
ref:opal,Seidl:Belle2025}, ranging from LEP measurements at the $Z^0$ boson resonance at
$\sqrt{S}=91.2\,\mathrm{GeV}$ to the newly 
included Belle data \cite{Seidl:Belle2025} just above the bottom threshold 
at $\sqrt{S}\simeq 10.5\,\mathrm{GeV}$, along with theoretical estimates at NLO accuracy 
based on our best fit (red solid lines), the $\kappa=1$ fit (dashed lines), 
AESSS (black solid lines), and NPC23 (dotted lines).

The only change in the selection of SIA data since the AESSS fit \cite{Aidala:2010bn} 
is the swap of the never published, preliminary BaBar data \cite{ref:babar} 
with the recent Belle measurement \cite{Seidl:Belle2025} at the same $\sqrt{S}$.
The latter data are systematically higher than the BaBar data used by AESSS which
explains why their $\eta$ FFs underestimate Belle data by roughly $15 - 20\%$.
The new fit, with and without scale modifications, as well as the NPC23 set
of FFs yield a good description of the suite of SIA data.
We note that the scale factors $\kappa(\sqrt{S})$ in SIA do not play a significant
role in obtaining a good fit but we include them for consistency.
To accommodate the new Belle data in our fit, the contributions of the different 
light and heavy quark flavors to the formation of eta mesons
are somewhat redistributed as compared to those found in the AESSS analysis 
as we shall illustrate next.

\begin{figure}[hbt!]
    \centering
    \includegraphics[width=0.9\linewidth]{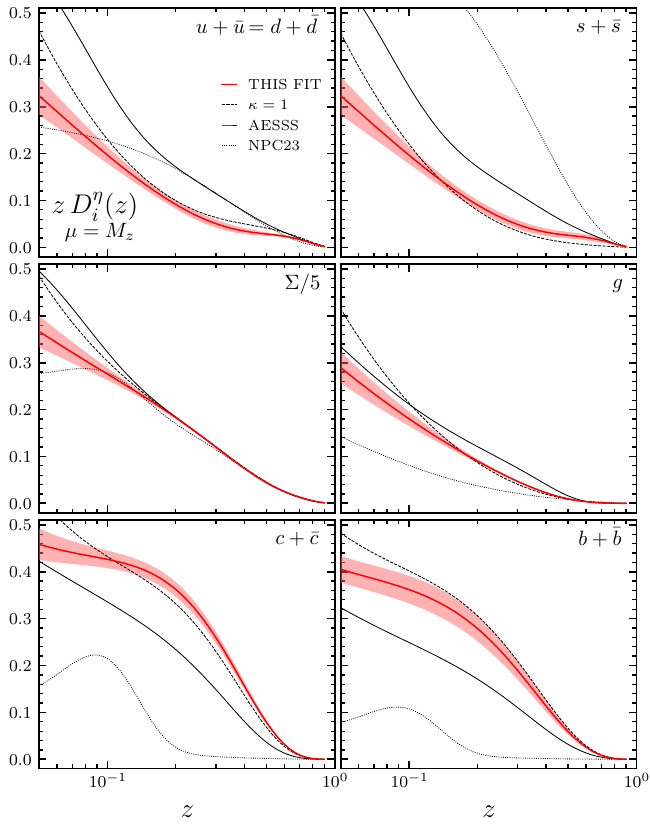}
    \caption{Parton-to-$\eta$ meson FFs $z\,D_i^{\eta}(z,\mu)$ (red lines) 
    as a function of momentum fraction $z$ for $i=u+\bar{u}=d+\bar{d}$. $s+\bar{s}$, 
    $c+\bar{c}$, $b+\bar{b}$, and $g$ at scale $\mu=M_Z$. 
    Also shown is the total quark singlet contribution $z\,\Sigma=\sum_{q=u,d,s,c,b} q+\bar{q}$
    but rescaled by a factor $1/5$.
    The shaded bands represent the $1$-$\sigma$ uncertainties as obtained from the 300 replicas
    of the best fit. For comparison, we also show the results of the $\kappa=1$ fit and
    for two other sets of FFs, AESSS \cite{Aidala:2010bn} and NPC23 \cite{Gao:2025bko}.}
    \label{fig:FFs}
\end{figure}  
In Figure~\ref{fig:FFs} we show the resulting $zD_i^{\eta}(z,\mu)$ of our best fit at 
scale $\mu=M_Z$ as a function of the momentum fraction $z$ along with corresponding 
uncertainties estimates at the $68\%$ C.L.\ (shaded bands) as obtained from the 300 replicas.
For comparison, we also show the results of the $\kappa=1$ fit as well as for
the sets of $\eta$ meson FFs from AESSS \cite{Aidala:2010bn} and NPC23 \cite{Gao:2025bko}.

Several observations are in order. Firstly, no matter what is assumed about the
separation of quark flavors in the various different fits, the total quark singlet contribution
$z\,D_{\Sigma}^{\eta}(z,M_Z)=\sum_{q=u,d,s,c,b} z[D^{\eta}_q(z,M_Z)+D^{\eta}_{\bar{q}}(z,M_Z)]$
turns out to be largely identical in the fitted region $z>0.1$ 
as it is tightly constrained by the precise SIA data at the $Z$ boson resonance from LEP. 
For the individual quark FFs, however, a very different picture emerges. 
As compared to the AESSS fit, noticeably smaller light quark FFs 
arise in our new fits (both with and without scale modification factors $\kappa(\sqrt{S})$) 
by redistributing some of their contributions to the heavy flavors such that the quark singlet
remains unchanged. In addition, the new gluon FF is also considerably smaller than in AESSS.

These findings can be traced back to the new suite of very precise hadroproduction data
for the first time included in our global analysis.
We recall, see Figs.~\ref{fig:mid} and \ref{fig:fwd_bwd}, that the AESSS fit grossly
overestimates most of the LHC data. 
Since $gg\to gg$ scattering is by far the most important partonic channel for
single inclusive particle production in $pp$ collisions
at midrapidity and small-to-medium values of $p_T$, 
the new fits first of all prefer a smaller gluon FF. 
At higher $p_T$ and for the forward rapidity data shown in Fig.~\ref{fig:fwd_bwd}, quark 
induced partonic processes play a more prominent role.
Due to the relative smallness of charm and bottom PDFs at 
the relevant scales and momentum fractions $x$, heavy quark fragmentation 
contributes relatively little to the hadroproduction results, hence also 
light quark FFs turn out to be reduced as compared to those found by AESSS. 
But this reduction needs to be compensated by larger charm and bottom FFs 
due to the well constrained total quark singlet FF.

Secondly, as can be inferred from Fig.~\ref{fig:FFs} as well, the FFs
obtained in the NPC23 analysis \cite{Gao:2025bko} of SIA and
$R_{\eta/\pi^{0}}$ data differ even more from our results than the AESSS set.
The contributions from different quark flavors are distributed in
a very different fashion. The strange quark FF is considerably larger than in
both our and the AESSS fits at the expense of much smaller heavy quark FFs.
Their fit also yields a rather small gluon FF. 
To some extent this differences are likely due to their higher cut of $p_T>4\,\mathrm{GeV}$
for $pp$ data and, as we have discussed, potential issues that arise when fitting
ratios of cross sections rather than cross sections. 
On the other hand, it certainly also illustrates potential shortcomings of current global
analyses of parton-to-$\eta$ meson FFs due to the complete lack of flavor tagged SIA
and SIDIS data, i.e., the fact that one has to rely not only on data but also
on assumptions about the flavor dependence of the $D_i^{\eta}(z,\mu)$. 
Hence, the uncertainty estimates may faithfully propagate experimental uncertainties 
to the extracted set of FFs but they do not account for all possible theoretical ambiguities.
One important source, which outnumbers the data driven uncertainties considerably, is the
scale dependence of the $pp$ cross sections, as was illustrated in 
Figs.~\ref{fig:mid} and \ref{fig:fwd_bwd}.

Finally, we note that despite starting with $D_{s+\bar{s}}^{\eta}<D_{u+\bar{u}}^{\eta}$ at our
input scale $\mu_0=1\,\mathrm{GeV}$, see Eqs.~(\ref{eq:quarks}) and (\ref{eq:ansatz}), 
the scale evolution to a large extent washes out
the input distributions, which are peaked at relatively large values of $z$, yielding
light quark FFs of almost equal size at $\mu=M_Z$ and that rise towards smaller values of $z$.
%
\begin{table}[tbh!]
\caption{\label{tab:exppiontab} Data sets, normalizations $\mathcal{N}_i$, 
and the partial and total $\chi^2$ values obtained in our best fit 
based on energy dependent scale factors $\kappa(\sqrt{S})$.}
\begin{tabular}{llcccc}
\hline\hline
experiment                   &       $\sqrt s$                               & data       & $\mathcal{N}_i$        & data   & $\chi^2$ \\
                             &                                               & type       &                        & fitted &          \\ \hline
PHENIX mid $y$              & $200 \text{ GeV}$                             & $2 \gamma$ \cite{Adler:2007phenix}  & 1.145  &  12    &   7.13   \\
                             &                                               & $3 \pi $ \cite{Adler:2007phenix}    & 1.069  &   6    &   4.31   \\
                             &                                               & 2006  \cite{Adare:2011phenix}       & 1.050  &  25    &  30.92   \\
                             & $510 \text{ GeV}$                             & 2025   \cite{Abdulameer:2025phenix} & 1.112  &  22    &   7.39   \\
PHENIX fwd $y$              & $200 \text{ GeV}$ \cite{Adare:2014phenix}     &                                     & 0.796  &  13    &   6.76   \\
                             & $500 \text{ GeV}$ \cite{Abdulameer:2025phenix}&                                     & 0.999  &  18    &  10.49   \\
ALICE                        & $2.76 \text{ TeV}$ \cite{Acharya:2017alice}   &                                     & 0.994  &   9    &   5.79   \\
                             & $7    \text{ TeV}$ \cite{Acharya:2021alice}   &                                     & 1.027  &  10    &   3.84   \\
                             & $8    \text{ TeV}$ \cite{Acharya:2018alice}   &                                     & 0.985  &  22    &  19.55   \\
                             & $13   \text{ TeV}$ \cite{Acharya:2021alice}   &                                     & 0.949  &  19    &  54.46   \\
LHCb \cite{Aaij:2024lhcb}    & $5.02 \text{ TeV}$                            &  very fwd                           & 1.021  &  21    &   8.76  \vspace{1mm} \\
                             &                                               &   fwd                               & 1.021  &  21    &  11.55   \\
                             & $13   \text{ TeV}$                            &  very fwd                           & 1.013  &  21    &   8.27   \\
                             &                                               &   fwd                               & 1.013  &  21    &  13.48   \\ \hline
{\bf PP data (sum)}          &                                               &                                     &        &  240   & 192.70   \\ \hline
BELLE \cite{Seidl:Belle2025} & $10.58 \text{ GeV}$                           &                                     & 1.015  &  30    &   9.04   \\                 
MARKII \cite{ref:mark2}      & $29 \text{ GeV}$                              &                                     & 0.983  &   7    &   3.69   \\
HRS  \cite{ref:hrs}          & $29 \text{ GeV}$                              &                                     & 0.556  &  13    &  26.38   \\
JADE 90  \cite{ref:jade2}    & $34.9 \text{ GeV}$                            &                                     &     -  &   3    &   0.43   \\
CELLO  \cite{ref:cello}      & $35 \text{ GeV}$                              &                                     & 0.998  &   4    &   1.35   \\
ALEPH92  \cite{ref:aleph1}   & $91.2 \text{ GeV}$                            &                                     &      - &   8    &   1.95   \\
ALEPH00  \cite{ref:aleph2}   & $91.2 \text{ GeV}$                            &                                     &      - &  18    &  14.63   \\
ALEPH02  \cite{ref:aleph3}   & $91.2 \text{ GeV}$                            &                                     &      - &   5    &  67.03   \\
L3 92  \cite{ref:l31}        & $91.2 \text{ GeV}$                            &                                     &      - &   3    &   6.32   \\
L3 94  \cite{ref:l32}        & $91.2 \text{ GeV}$                            &                                     &      - &   8    &  14.48   \\
OPAL   \cite{ref:opal}       & $91.2 \text{ GeV}$                            &                                     &      - &   9    &  12.40   \\
JADE85  \cite{ref:jade1}     & $34.4 \text{ GeV}$                            &                                     &      - &   1    &  12.30   \\ \hline
{\bf SIA data (sum)}         &                                               &                                     &        &  109   & 170.00   \\ \hline
 \hline                                    
{\bf TOTAL:}                 &                                               &                                     &        &  349   & 362.70    \\
\hline\hline
\end{tabular}
\end{table} 

To conclude this section, we summarize in Table~\ref{tab:exppiontab}
our updated global QCD analysis of 
parton-to-$\eta$ meson FFs by quoting 
the partial contributions for each set of data to the total $\chi^2$
of our best fit based on energy dependent scale factors $\kappa(\sqrt{S})$.
Also shown are the individual number of data points utilized in 
the fit, i.e., that pass our cut of $p_T>1.5\,\mathrm{GeV}$ and $z>0.1$ in
case of $pp$ and SIA data, respectively.
The optimum normalization shifts $\mathcal{N}_i$ are, as usual, 
determined analytically as specified in Eq.~(6) of 
ref.~\cite{deFlorian:2014xna}, i.e., they are no free parameters in
the fit.

As we have shown in detail already in Figs.~\ref{fig:mid} - \ref{fig:SIA}
and can be inferred also from Tab.~\ref{tab:exppiontab},
the overall description of the experimental results by theoretical 
estimates at NLO accuracy based on our new set of FFs is very good.
The total $\chi^2$ of 170 units for the 109 SIA data points is mainly
due to the rather poor description of the 5 points from the 2002 
ALEPH measurement \cite{ref:aleph3} and can be viewed as an outlier.

We also wish to mention that the total $\chi^2$ of the $\kappa=1$ fit is
about 100 units worse than for the best fit with the difference mainly stemming
from the PHENIX $pp$ data as can be anticipated from 
Figs.~\ref{fig:mid} and \ref{fig:fwd_bwd}.
The SIA data are equally well described. This fit, with the more
straightforward choice of the default scale $p_T$ in the computation
of hadroproduction cross sections, is a good alternative to our best fit as long as one
is only interested in phenomenology at LHC energies. Both new fits, with
$\kappa(\sqrt{S})$ and $\kappa=1$, as well as
the 300 replicas for our best fit are publicly available 
in LHAPDF format \cite{ref:github}.

\section{Summary and Outlook}
We have updated the extraction of nonperturbative parton-to-eta meson FFs by means of a global
QCD analysis at NLO accuracy based on the latest suite of hadroproduction data 
collected at the CERN-LHC and BNL-RHIC in a vast range of c.m.s.\ energies and the
well-known measurements in single inclusive electron-positron annihilation 
supplemented by recent BELLE data.

To reconcile data taken at the lowest and the highest c.m.s.\ energies we exploited the
freedom in the choice of theoretical renormalization and factorization scales.
To this end, we successfully adopted the same energy dependent scales that were introduced in 
corresponding fits of FFs for pion, kaons, and unidentified charged hadrons. 
An alternative fit based on the conventional choice of scales could only reproduce
LHC data but led to tensions with PHENIX data at lower c.m.s.\ energies.

As compared to a previous fit of eta meson FFs, solely based on SIA and
a single set of RHIC data, that is found to grossly overestimate all LHC data, 
the huge amount of recent hadroproduction data now allows us to constrain 
the gluon fragmentation function much more precisely and with a 
more elaborate functional form. 
The availability of new hadroproduction data, not only in a vast range of
c.m.s. energies but also for both mid and forward rapidities of the
observed $\eta$ meson, helped to better determine also the (anti)quark-to-$\eta$ meson
FFs by complementing the information from SIA data. The latter yield a strong
constraint of the total quark singlet fragmentation function at the $Z$ boson resonance.
In combination with hadroproduction data and, indirectly, 
through evolution effects, they provide at least some discriminating power
between contributions from light, strange, and heavy flavors
beyond what was available to previous fits.

The flavor separation for eta meson FFs, however, still relies on considerably more 
assumptions than in the case of pion or kaons, and obtained uncertainty estimates have to
be taken with caution. This is because of the complete lack of 
flavor-tagged measurements in SIA, and, most importantly, experimental information
from SIDIS with detected $\eta$ mesons.
Considerable progress in the latter case can be expected from the forthcoming first
Electron-Ion Collider \cite{Accardi:2012qut,AbdulKhalek:2021gbh} to be constructed at BNL.

On the theoretical side, on can hope to see first truly global analyses 
at NNLO accuracy in the near future thanks to the recent major progress on
calculations for SIDIS \cite{Goyal:2023zdi,Bonino:2024qbh}
and single inclusive hadron production in $pp$ collisions \cite{Czakon:2025yti}.
Ultimately, this will hopefully ease all the currently known tensions between
data collected at different ends of the vast range of c.m.s.\ energies
probed experimentally, and will render the implementation of energy dependent scale
factors unnecessary.\\

\section*{Acknowledgments}
%
We thank Ralf Seidl and Joshua Koenig for their help with BELLE and ALICE data, respectively, 
and also Jun Gao and ChongYang Liu for providing us with the NPC23 sets. 
This work was supported in part by CONICET, ANPCyT, UBACyT, and
the Bundesministerium f{\"u}r Bildung und Forschung (BMBF) under grant no.\ 05P21VTCAA. 
%

%
\end{document}